\documentclass[12pt]{article}
\usepackage{amssymb}

\makeatletter
\def\numberbysection{\@addtoreset{equation}{section}
 	\def\theequation{\thesection.\arabic{equation}}}
\makeatother

\numberbysection

\newcommand{\be}{\begin{eqnarray}}
\newcommand{\ee}{\end{eqnarray}}
\newcommand{\non}{\nonumber}
\newcommand{\tr}{\mathop{\rm tr}\nolimits}
\newcommand{\id}{\mathbb{I}}
\newcommand{\csch}{\mathop{\rm csch}\nolimits}
\newcommand{\sech}{\mathop{\rm sech}\nolimits}

\begin{document}

\begin{titlepage}
\strut\hfill UMTG--246
\vspace{.5in}
\begin{center}

\LARGE Bethe Ansatz derived from the functional relations \\
\LARGE of the open XXZ chain for new special cases \\[1.0in]
\large Rajan Murgan and Rafael I. Nepomechie\\[0.8in]
\large Physics Department, P.O. Box 248046, University of Miami\\[0.2in]  
\large Coral Gables, FL 33124 USA\\

\end{center}

\vspace{.5in}

\begin{abstract}
The transfer matrix of the general integrable open XXZ quantum spin
chain obeys certain functional relations at roots of unity.  By
exploiting these functional relations, we determine the Bethe Ansatz
solution for the transfer matrix eigenvalues for the special cases
that all but one of the boundary parameters are zero, and the bulk
anisotropy parameter is $i \pi/3\,, i \pi/5 \,, \ldots$.
\end{abstract}
\end{titlepage}

\setcounter{footnote}{0}

\section{Introduction}\label{sec:intro}

The open XXZ quantum spin chain with general integrable boundary terms
\cite{dVGR} is a fundamental integrable model with boundary, which has
applications in condensed matter physics, statistical mechanics and
string theory.  Although this model remains unsolved, the special case
of diagonal boundary terms was solved long ago \cite{Ga, ABBBQ, Sk},
and some progress on the more general case has been achieved recently
by two different approaches.  One approach, pursued by Cao {\it et
al.} \cite{CLSW} is an adaptation of the generalized algebraic Bethe
Ansatz \cite{Ba, FT} to open chains.  Another approach, which was
developed in \cite{XX}-\cite{NR} and which we pursue further here,
exploits the functional relations obeyed by the transfer matrix at
roots of unity.  It is based on fusion \cite{fusion}, the truncation
of the fusion hierarchy at roots of unity \cite{truncation}, and the
Bazhanov-Reshetikhin \cite{BR} solution of RSOS models.

Both approaches lead to a Bethe Ansatz solution for the special case
that the boundary parameters obey a certain constraint. Namely,
(following the notation of the second reference in \cite{Ne} where
$\alpha_{-}\,, \beta_{-}\,,
\theta_{-}$ and $\alpha_{+}\,, \beta_{+}\,,
\theta_{+}$ denote the left and right boundary parameters,
respectively, and $N$ is the number of spins in the chain),
\be
\alpha_{-} + \beta_{-} + \alpha_{+} + \beta_{+} = \pm (\theta_{-} - 
\theta_{+}) + \eta k \,,
\label{constraint}
\ee
where $k$ is an even integer if $N$ is odd, and is an odd integer if
$N$ is even.
This solution has been used to derive a nonlinear integral equation
for the sine-Gordon model on an interval \cite{NLIE}, and has been
generalized to other models \cite{generalization}.

Despite these successes, it would be desirable to find the solution
for general values of the boundary parameters; i.e., when the
constraint (\ref{constraint}) is not satisfied.
In the functional relation approach,  the main difficulty lies in
recasting the functional relations (which are known \cite{XXZ, Ne} for
general values of the boundary parameters) as the condition that a 
certain determinant vanish. In this note we report the solution of
this problem (and hence, the Bethe Ansatz expression for the transfer 
matrix eigenvalues) for the special cases that all but one of the
boundary parameters are zero, and the bulk anisotropy has values 
$\eta = {i \pi\over 3}\,, {i \pi\over 5}\,, \ldots $. It may be
possible to extend this analysis to more general cases.

In Section \ref{sec:transfer}, we briefly review the construction of
the transfer matrix and the functional relations which it satisfies at
roots of unity.  In Section \ref{sec:BA} we present our main results;
namely, the Bethe Ansatz solution for the transfer matrix eigenvalues
when all but one of the boundary parameters vanish.  We conclude in
Section \ref{sec:discuss} with a brief discussion of these results.
In an Appendix we briefly review the solution \cite{Ne, NR} for the
case that the constraint (\ref{constraint}) is satisfied, in order to
facilitate comparison with the new cases considered here.

\section{Transfer matrix and functional relations}\label{sec:transfer}

The transfer matrix $t(u)$ of the open XXZ chain with general integrable
boundary terms is given by \cite{Sk}
\be
t(u) = \tr_{0} K^{+}_{0}(u)\  
T_{0}(u)\  K^{-}_{0}(u)\ \hat T_{0}(u)\,,
\label{transfer}
\ee
where $T_{0}(u)$ and $\hat T_{0}(u)$ are the monodromy matrices 
\be
T_{0}(u) = R_{0N}(u) \cdots  R_{01}(u) \,,  \qquad 
\hat T_{0}(u) = R_{01}(u) \cdots  R_{0N}(u) \,,
\label{monodromy}
\ee
and $\tr_{0}$ denotes trace over the ``auxiliary space'' 0.
The $R$ matrix is given by
\be
R(u) = \left( \begin{array}{cccc}
	\sinh  (u + \eta) &0            &0           &0            \\
	0                 &\sinh  u     &\sinh \eta  &0            \\
	0                 &\sinh \eta   &\sinh  u    &0            \\
	0                 &0            &0           &\sinh  (u + \eta)
\end{array} \right) \,,
\label{bulkRmatrix}
\ee 
where $\eta$ is the bulk anisotropy parameter; and $K^{\mp}(u)$ are
$2 \times 2$ matrices whose components
are given by \cite{dVGR, GZ}
\be
K_{11}^{-}(u) &=& 2 \left( \sinh \alpha_{-} \cosh \beta_{-} \cosh u +
\cosh \alpha_{-} \sinh \beta_{-} \sinh u \right) \non \\
K_{22}^{-}(u) &=& 2 \left( \sinh \alpha_{-} \cosh \beta_{-} \cosh u -
\cosh \alpha_{-} \sinh \beta_{-} \sinh u \right) \non \\
K_{12}^{-}(u) &=& e^{\theta_{-}} \sinh  2u \,, \qquad 
K_{21}^{-}(u) = e^{-\theta_{-}} \sinh  2u \,,
\label{Kminuscomponents}
\ee
and
\be
K_{11}^{+}(u) &=& -2 \left( \sinh \alpha_{+} \cosh \beta_{+} \cosh (u+\eta) 
- \cosh \alpha_{+} \sinh \beta_{+} \sinh (u+\eta) \right) \non \\
K_{22}^{+}(u) &=& -2 \left( \sinh \alpha_{+} \cosh \beta_{+} \cosh (u+\eta) 
+ \cosh \alpha_{+} \sinh \beta_{+} \sinh (u+\eta) \right) \non \\
K_{12}^{+}(u) &=& -e^{\theta_{+}} \sinh  2(u+\eta) \,, \qquad 
K_{21}^{+}(u) = -e^{-\theta_{+}} \sinh  2(u+\eta) \,,
\label{Kpluscomponents}
\ee
where $\alpha_{\mp} \,, \beta_{\mp} \,, \theta_{\mp}$ are the boundary
parameters. \footnote{Following \cite{Ne, NR}, we use a
parametrization of the boundary parameters which differs from that in 
\cite{dVGR, GZ}. Specifically, the matrices $K^{\mp}(u)$ are equal to those 
appearing in the second reference in \cite{Ne} divided by the factors 
$\kappa_{\mp}$, respectively.}  

In addition to the fundamental commutativity property
\be
\left[ t(u)\,, t(v) \right] = 0  \,,
\label{commutativity}
\ee 
the transfer matrix also has $i \pi$ periodicity
\be
t(u+ i \pi) = t(u) \,,
\label{periodicity}
\ee
crossing symmetry
\be
t(-u - \eta)= t(u) \,,
\label{transfercrossing}
\ee
and the asymptotic behavior 
\be
t(u) \sim -\cosh(\theta_{-}-\theta_{+})
{e^{u(2N+4)+\eta (N+2)}\over 2^{2N+1}} \id + 
\ldots \qquad \mbox{for} \qquad
u\rightarrow \infty \,.
\label{transfasympt}
\ee

For bulk anisotropy values $\eta = {i \pi\over p+1}$, with 
$p= 1 \,, 2 \,, \ldots $, the transfer matrix obeys functional
relations of order $p+1$  \cite{XXZ, Ne}
\be
\lefteqn{t(u) t(u +\eta) \ldots t(u + p \eta)} \non \\
&-& \delta (u-\eta) t(u +\eta) t(u +2\eta) 
\ldots t(u + (p-1)\eta) \non \\
&-& \delta (u) t(u +2\eta) t(u +3\eta)
\ldots t(u + p \eta) \non \\
&-& \delta (u+\eta) t(u) t(u +3\eta) t(u +4\eta) 
\ldots t(u + p \eta) \non \\
&-& \delta (u+2\eta) t(u) t(u +\eta) t(u +4\eta) 
\ldots t(u + p \eta) - \ldots \non \\
&-& \delta (u+(p-1)\eta) t(u) t(u +\eta) 
\ldots t(u +  (p-2)\eta) \non \\
&+& \ldots  = f(u) \,.
\label{funcrltn}
\ee 
For example, for the case $p=2$, the functional relation is
\be
t(u) t(u+\eta) t(u+2\eta) 
- \delta(u-\eta) t(u+\eta) 
- \delta(u) t(u+2\eta) 
- \delta(u+\eta) t(u) 
= f(u) \,.
\ee
The functions $\delta(u)$ and $f(u)$ are given in terms of the
boundary parameters  $\alpha_{\mp} \,, \beta_{\mp} \,, \theta_{\mp}$
by
\be
\delta(u) = \delta_{0}(u) \delta_{1}(u) \,, \qquad 
f(u) = f_{0}(u) f_{1}(u) \,,
\ee
where
\be
\delta_{0}(u) &=& \left( \sinh u \sinh(u + 2\eta) \right)^{2N} {\sinh 2u
\sinh (2u + 4\eta)\over \sinh(2u+\eta) \sinh(2u+3\eta)}\,, \label{delta0} \\
\delta_{1}(u) &=&  2^{4} \sinh(u + \eta + \alpha_{-}) \sinh(u + \eta - \alpha_{-})
\cosh(u + \eta + \beta_{-}) \cosh(u + \eta - \beta_{-})  \non \\
& \times & \sinh(u + \eta + \alpha_{+}) \sinh(u + \eta - \alpha_{+})
\cosh(u + \eta + \beta_{+}) \cosh(u + \eta - \beta_{+}) \,,
\label{delta1}
\ee
and therefore, 
\be
\delta(u+ i\pi) =\delta(u) \,, \qquad \delta(-u -2\eta) =\delta(u)
\,.
\ee 
For $p$ even,
\be
f_{0}(u) &=& (-1)^{N+1} 2^{-2 p N} \sinh^{2N} \left( (p+1)u \right)
\,, \label{f0} \\
f_{1}(u) &=& (-1)^{N+1} 2^{3-2 p} \Big( \non \\
& & \hspace{-0.2in}
\sinh \left( (p+1) \alpha_{-} \right)\cosh \left( (p+1) \beta_{-} \right)
\sinh \left( (p+1) \alpha_{+} \right)\cosh \left( (p+1) \beta_{+} \right)
\cosh^{2} \left( (p+1)u \right) \non \\
&-&
\cosh \left( (p+1) \alpha_{-} \right)\sinh \left( (p+1) \beta_{-} \right)
\cosh \left( (p+1) \alpha_{+} \right)\sinh \left( (p+1) \beta_{+} \right)
\sinh^{2} \left( (p+1)u \right) \non \\
&-&
(-1)^{N} \cosh \left( (p+1)(\theta_{-}-\theta_{+}) \right)
\sinh^{2} \left( (p+1)u \right) \cosh^{2} \left( (p+1)u \right) 
\Big) \,.
\label{f1}
\ee
For $p$ odd,
\be
f_{0}(u) &=& (-1)^{N+1} 2^{-2 p N} \sinh^{2N} \left( (p+1)u \right)
\tanh^{2} \left( (p+1)u \right)
\,, \label{f0odd} \\
f_{1}(u) &=& -2^{3-2 p} \Big( \non \\
& & \hspace{-0.2in}
\cosh \left( (p+1) \alpha_{-} \right)\cosh \left( (p+1) \beta_{-} \right)
\cosh \left( (p+1) \alpha_{+} \right)\cosh \left( (p+1) \beta_{+} \right)
\sinh^{2} \left( (p+1)u \right) \non \\
&-&
\sinh \left( (p+1) \alpha_{-} \right)\sinh \left( (p+1) \beta_{-} \right)
\sinh \left( (p+1) \alpha_{+} \right)\sinh \left( (p+1) \beta_{+} \right)
\cosh^{2} \left( (p+1)u \right) \non \\
&+&
(-1)^{N} \cosh \left( (p+1)(\theta_{-}-\theta_{+}) \right)
\sinh^{2} \left( (p+1)u \right) \cosh^{2} \left( (p+1)u \right) 
\Big) \,. \label{f1odd}
\ee 
Hence, $f(u)$ satisfies
\be
f(u + \eta) = f(u) \,, \qquad f(-u)=f(u) \,.
\ee

The commutativity property (\ref{commutativity}) implies that the
eigenvectors $|\Lambda\rangle$ of the transfer matrix $t(u)$ are
independent of the spectral parameter $u$.  Hence, the corresponding
eigenvalues $\Lambda(u)$ obey the same functional relations
(\ref{funcrltn}), as well as the properties (\ref{periodicity}) -
(\ref{transfasympt}).

\section{Bethe Ansatz solution for new special cases}\label{sec:BA}

We henceforth restrict to {\it even} values of $p$ (i.e., bulk
anisotropy values $\eta = {i \pi\over 3}\,, {i \pi\over 5}\,, \ldots
$), and consider the various special cases that all but one of the
boundary parameters are zero.

\subsection{$\alpha_{-} \ne 0$}\label{subsec:alpham}

For the case that all boundary parameters are zero except for
$\alpha_{-}$ (or, similarly, $\alpha_{+}$), we find that the
functional relations (\ref{funcrltn}) for the transfer matrix
eigenvalues can be written as
\be
\det {\cal M} = 0 \,,
\label{det}
\ee
where ${\cal M}$ is given by the $(p+1) \times (p+1)$ matrix
\be
{\cal M} = \left(
\begin{array}{cccccccc}
    \Lambda(u) & -h(u) & 0  & \ldots  & 0 & -h(-u+p \eta)  \\
    -h(-u) & \Lambda(u+p\eta) & -h(u+p \eta)  & \ldots  & 0 & 0  \\
    \vdots  & \vdots & \vdots & \ddots 
    & \vdots  & \vdots    \\
   -h(u+p^{2} \eta)  & 0 & 0 & \ldots  & -h(-u-p(p-1) \eta) &
    \Lambda(u+p^{2}\eta)
\end{array} \right)  \,,
\label{calMalpha}
\ee
(whose successive rows are obtained by simultaneously shifting $u \mapsto u+ p \eta$
and cyclically permuting the columns to the right)
provided that there exists a function $h(u)$ which has the properties
\be
h(u + 2 i \pi) = h \left(u +2(p+1)\eta \right) &=& h(u) \,, \label{cond0} \\
h(u+(p+2)\eta)\ h(-u-(p+2)\eta) &=& \delta(u) \,, \label{cond1} \\
\prod_{j=0}^{p} h(u+2j\eta) + \prod_{j=0}^{p} h(-u-2j\eta) &=& f(u) 
\,. \label{cond2} 
\ee

To solve for $h(u)$, we set 
\be
h(u) = h_{0}(u) h_{1}(u) \,,
\label{h}
\ee
with 
\be
h_{0}(u) = (-1)^{N} \sinh^{2N}(u+\eta){\sinh(2u+2\eta)\over \sinh(2u+\eta)}
\,.
\label{h0}
\ee
Noting that 
\be
h_{0}(u+(p+2)\eta)\ h_{0}(-u-(p+2)\eta) &=& \delta_{0}(u) \,, \non \\
\prod_{j=0}^{p} h_{0}(u+2j\eta) = \prod_{j=0}^{p} h_{0}(-u-2j\eta) &=&
f_{0}(u)  \,,
\ee
where $\delta_{0}(u)$ and $f_{0}(u)$ are given by (\ref{delta0}) and
(\ref{f0}), respectively, we see that $h_{1}(u)$ must satisfy
\be
h_{1}(u+(p+2)\eta)\ h_{1}(-u-(p+2)\eta) &=& \delta_{1}(u) \,, \label{h1cond1} \\
\prod_{j=0}^{p} h_{1}(u+2j\eta) + \prod_{j=0}^{p} h_{1}(-u-2j\eta) &=&
f_{1}(u)  \,. \label{h1cond2}
\ee
Eliminating $h_{1}(-u-2j\eta)$ in (\ref{h1cond2}) using
(\ref{h1cond1}), we obtain 
\be
z(u)^{2} -z(u) f_{1}(u) +  \prod_{j=0}^{p} \delta_{1}\left(u+(2j-1)\eta\right) = 0 \,,
\label{quadratic}
\ee
where 
\be
z(u) = \prod_{j=0}^{p} h_{1}(u+2j\eta) \,.
\label{producth1}
\ee
Solving the quadratic equation (\ref{quadratic}) for $z(u)$, making use of the
explicit expressions (\ref{delta1}) and (\ref{f1}) for $\delta_{1}(u)$
and $f_{1}(u)$, respectively, we obtain
\be
z(u)=2^{-2(p-1)} \cosh^{2}\left( (p+1)u \right) \sinh \left( (p+1)u \right)
\left( \sinh \left( (p+1)u \right) \pm \sinh \left( (p+1) \alpha_{-} \right)
\right) \,.
\ee
Notice that this expression for $z(u)$ has periodicity $2\eta$, which
is consistent with (\ref{producth1}) and the assumed periodicity
(\ref{cond0}). Corresponding solutions of (\ref{producth1}) for $h_{1}(u)$ are 
\be
h_{1}(u) = -4 \cosh^{2}u \sinh u \sinh(u \mp \alpha_{-})
{\cosh\left({1\over 2}(u \pm \alpha_{-}+\eta) \right)\over 
 \cosh\left({1\over 2}(u \mp \alpha_{-}-\eta) \right)}
\,. \label{h1}
\ee
In short, a function $h(u)$ which satisfies (\ref{cond0}) - (\ref{cond2})
is given by 
\be
h(u) &=& (-1)^{N+1} 4\sinh^{2N}(u+\eta){\sinh(2u+2\eta)\over \sinh(2u+\eta)}
\cosh^{2}u \sinh u  \non \\
&\times& \sinh(u-\alpha_{-}) {\cosh\left({1\over 2}(u+\alpha_{-}+\eta) \right)\over 
 \cosh\left({1\over 2}(u-\alpha_{-}-\eta) \right)} \,.
\label{halpha}
\ee

The structure of the matrix ${\cal M}$ (\ref{calMalpha}) suggests that its 
null eigenvector has the form $\big( Q(u)\,, Q(u+p\eta) \,, \ldots
\,, Q(u+p^{2}\eta) \big)$, where $Q(u)$ has the periodicity property
\be
Q(u + 2i\pi) = Q(u) \,.
\label{Qperiodicity}
\ee
It follows that the transfer matrix
eigenvalues are given by
\be
\Lambda(u) = h(u) {Q(u + p\eta)\over Q(u)} 
+ h(-u+p \eta) {Q(u -p\eta)\over Q(u)}  \,,
\label{eigenvalues} 
\ee
which evidently has the form of Baxter's $TQ$ relation.
We make the Ansatz
\be
Q(u) = \prod_{j=1}^{M} 
\sinh \left( {1\over 2}(u - u_{j}) \right)
\sinh \left( {1\over 2}(u + u_{j} - p\eta) \right) \,,
\label{Q}
\ee 
which has the periodicity (\ref{Qperiodicity}) as well as the crossing
property \footnote{Note that
$\Lambda(u) = \Lambda(-u+ p\eta) = \Lambda(-u -\eta)$, where the
first equality follows
from (\ref{eigenvalues}) and (\ref{Qcrossing}), and the second
equality follows from the $i \pi$ periodicity of $\Lambda(u)$ (which,
however, is not manifest from (\ref{eigenvalues}).)}
\be
Q(-u+ p\eta) = Q(u) \,.
\label{Qcrossing}
\ee
The asymptotic behavior (\ref{transfasympt}) is consistent with having
$M$ (the number of zeros $u_{j}$ of $Q(u)$) given by
\be
M=N+p+1
\,, \label{M}
\ee
which we have confirmed numerically for small values of $N$ and $p$.
Analyticity of $\Lambda(u)$ implies the Bethe Ansatz equations
\be
{h(u_{j})\over h(-u_{j}+p\eta)} = 
-{Q(u_{j}-p\eta)\over Q(u_{j}+p\eta)} \,, 
\qquad j = 1 \,, \ldots \,, M \,.
\label{BAeqs}
\ee

To summarize, for the special case that $p$ is even and all boundary
parameters are zero except for $\alpha_{-}$, the eigenvalues of the
transfer matrix (\ref{transfer}) are given by (\ref{eigenvalues}),
where $h(u)$ is given by (\ref{halpha}), and $Q(u)$ is given by
(\ref{Q}),(\ref{M}), with zeros $u_{j}$ given by (\ref{BAeqs}).

We observe that for the special case that we are considering, the
corresponding Hamiltonian is {\it not} of the usual XXZ form.  Indeed,
$t'(0)$ (the first derivative of the transfer matrix evaluated at
$u=0$) is proportional to $\sigma^{x}_{N}$.  Hence, to obtain a
nontrivial integrable Hamiltonian, one must consider the second
derivative of the transfer matrix. We find
\be
t''(0) &=& -16 \sinh^{2N-1} \eta \cosh \eta \sinh \alpha_{-}
\Bigg(  \left\{ 
\sigma^{x}_{N} \,, \sum_{n=1}^{N-1} H_{n\,, n+1} \right\} \non \\
&+& (N \cosh \eta + \sinh \eta \tanh \eta)  \sigma^{x}_{N}
+ {\sinh \eta\over \sinh \alpha_{-}} \sigma^{x}_{1} \sigma^{x}_{N} 
\Bigg)
\,,
\ee 
where $H_{n\,, n+1}$ is given by
\be
H_{n\,, n+1} = {1\over 2}\left( \sigma^{x}_{n}\sigma^{x}_{n+1}
+\sigma^{y}_{n}\sigma^{y}_{n+1}+\cosh \eta\ \sigma^{z}_{n}\sigma^{z}_{n+1}
\right) \,.
\ee

\subsection{$\beta_{-} \ne 0$}\label{subsec:betam}

For the case that all boundary parameters are zero except for
$\beta_{-}$ (or, similarly, $\beta_{+}$), we find that the
functional relations (\ref{funcrltn}) for the transfer matrix
eigenvalues can again be written in the form (\ref{det}),
where now the matrix ${\cal M}$ is given by 
\be
{\cal M} = \left(
\begin{array}{cccccccc}
    \Lambda(u) & -h(u) & 0  & \ldots  & 0 & -h(-u- \eta)  \\
    -h(-u-(p+1)\eta) & \Lambda(u+p\eta) & -h(u+p \eta)  & \ldots  & 0 & 0  \\
    \vdots  & \vdots & \vdots & \ddots 
    & \vdots  & \vdots    \\
   -h(u+p^{2} \eta)  & 0 & 0 & \ldots  & -h(-u-(p^{2}+1) \eta) &
    \Lambda(u+p^{2}\eta)
\end{array} \right)  \,,
\label{calMbeta}
\ee
if $h(u)$ satisfies
\be
h(u + 2 i \pi) = h \left(u +2(p+1)\eta \right) &=& h(u) \,,
\label{cond0beta} \\
h(u+(p+2)\eta)\ h(-u-\eta) &=& \delta(u) \,, \label{cond1beta} \\
\prod_{j=0}^{p} h(u+2j\eta) + \prod_{j=0}^{p} h(-u-(2j+1)\eta) &=& f(u) 
\,. \label{cond2beta} 
\ee

Proceeding similarly to the previous case, we now find 
\be
h(u) &=& (-1)^{N} 4\sinh^{2N}(u+\eta){\sinh(2u+2\eta)\over \sinh(2u+\eta)}
\sinh^{2}u \cosh u \left(\cosh u + (-1)^{p\over 2} i \sinh \beta_{-} \right) \,.
\non \\
\label{hbeta}
\ee
The transfer matrix eigenvalues are now given by
\be
\Lambda(u) = h(u) {Q(u + p\eta)\over Q(u)} 
+ h(-u - \eta) {Q(u -p\eta)\over Q(u)}  \,,
\label{eigenvaluesbeta} 
\ee
with
\be
Q(u) = \prod_{j=1}^{M} 
\sinh \left( {1\over 2}(u - u_{j}) \right)
\sinh \left( {1\over 2}(u + u_{j} + \eta) \right) \,,
\label{Qbeta}
\ee 
which satisfies $Q(u + 2i\pi) = Q(u)$ and $Q(-u-\eta) = Q(u)$; and
\be
M=N+p
\,. \label{Mbeta}
\ee
Moreover, the Bethe Ansatz equations for the zeros $u_{j}$ take the form 
\be
{h(u_{j})\over h(-u_{j}-\eta)} = 
-{Q(u_{j}-p\eta)\over Q(u_{j}+p\eta)} \,, 
\qquad j = 1 \,, \ldots \,, M \,.
\label{BAeqsbeta}
\ee

For this case, $t'(0)=0$, and 
\be
t''(0)= -16 \cosh \eta \sinh^{2N}\eta \left( \sigma^{x}_{1}+\sinh 
\beta_{-}\ \sigma^{z}_{1} \right) \sigma^{x}_{N}
\,.
\ee 
Higher derivatives yield more complicated expressions.

\subsection{$\theta_{\mp} \ne 0$}\label{subsec:thetamp}

For the case that all boundary parameters are zero except for
$\theta_{-}$ and $\theta_{+}$ (quantities of interest depend only 
on the difference $\theta_{-}-\theta_{+}$), we find that the
functional relations (\ref{funcrltn}) for the transfer matrix
eigenvalues can be written in the form (\ref{det}),
where the matrix ${\cal M}$ is given by 
\be
{\cal M} = \left(
\begin{array}{cccccccc}
    \Lambda(u) & -h^{(2)}(-u-\eta) & 0  & \ldots  & 0 & -h^{(1)}(u)  \\
    -h^{(1)}(u+\eta) & \Lambda(u+\eta) & -h^{(2)}(-u-2\eta)  & \ldots  & 0 & 0  \\
    \vdots  & \vdots & \vdots & \ddots 
    & \vdots  & \vdots    \\
   -h^{(2)}(-u-(p+1)\eta)  & 0 & 0 & \ldots  & -h^{(1)}(u+p\eta) &
    \Lambda(u+p\eta)
\end{array} \right)  \,,
\label{calMtheta}
\ee
(whose successive rows are obtained by simultaneously shifting $u \mapsto u+ \eta$
and cyclically permuting the columns to the right), if 
the functions $h^{(1)}(u)$ and $h^{(2)}(u)$ satisfy
\be
h^{(k)}(u + i \pi) = h^{(k)} \left(u + (p+1)\eta \right) &=& h^{(k)}(u) \,,
\quad k = 1\,, 2 \,, 
\label{cond0theta} \\
h^{(1)}(u+ \eta)\ h^{(2)}(-u-\eta) &=& \delta(u) \,, \label{cond1theta} \\
\prod_{j=0}^{p} h^{(1)}(u+j\eta) + \prod_{j=0}^{p} h^{(2)}(-u-j\eta) &=& f(u) 
\,. \label{cond2theta} 
\ee

We find
\be
h^{(1)}(u) &=& (-1)^{N} e^{\theta_{+}-\theta_{-}}\sinh^{2N}(u+\eta)
{\sinh(2u+2\eta)\over \sinh(2u+\eta)}
\sinh^{2}2u \,,\non \\
h^{(2)}(u) &=& (-1)^{N} e^{\theta_{-}-\theta_{+}}\sinh^{2N}(u+\eta)
{\sinh(2u+2\eta)\over \sinh(2u+\eta)}
\sinh^{2}2u \,.
\label{htheta}
\ee
The transfer matrix eigenvalues are given by
\be
\Lambda(u) = h^{(1)}(u) {Q(u -\eta)\over Q(u)} 
+ h^{(2)}(-u - \eta) {Q(u +\eta)\over Q(u)}  \,,
\label{eigenvaluestheta} 
\ee
with, for $N$ even,
\be
Q(u) = \prod_{j=1}^{2M} 
\sinh (u - u_{j}) \,,
\label{Qtheta}
\ee 
which satisfies $Q(u + i\pi) = Q(u)$; and 
\be
M={1\over 2}(N+p)
\,. \label{Mtheta}
\ee
The Bethe Ansatz equations for the zeros $u_{j}$ take the form 
\be
{h^{(1)}(u_{j})\over h^{(2)}(-u_{j}-\eta)} = 
-{Q(u_{j}+\eta)\over Q(u_{j}-\eta)} \,, 
\qquad j = 1 \,, \ldots \,, M \,.
\label{BAeqstheta}
\ee

For this case, also $t'(0)=0$, and 
\be
t''(0) &=& -16 \cosh \eta \sinh^{2N}\eta \Big( 
\cosh \theta_{-} \cosh \theta_{+}\ \sigma^{x}_{1} \sigma^{x}_{N} 
+ i \cosh \theta_{-} \sinh \theta_{+}\ \sigma^{x}_{1} \sigma^{y}_{N} 
\non \\ 
&+& i \sinh \theta_{-} \cosh \theta_{+}\ \sigma^{y}_{1} \sigma^{x}_{N} 
- \sinh \theta_{-} \sinh \theta_{+}\ \sigma^{y}_{1} \sigma^{y}_{N} 
\Big) \,.
\ee 

\section{Discussion}\label{sec:discuss}

We have checked these solutions numerically for chains of length up to
$N=6$, and have verified that they give the complete set of $2^{N}$
eigenvalues. Hence, completeness is achieved more simply than in the 
case that the constraint (\ref{constraint}) is satisfied \cite{NR}.

We emphasize that, in contrast to the solution for the case that the
constraint (\ref{constraint}) is satisfied, these solutions do {\it
not} hold for generic values of the bulk anisotropy.  Indeed, these
solutions hold only for $\eta = {i \pi\over 3}\,, {i \pi\over 5}\,,
\ldots $. 
Also, while the $Q(u)$ functions have periodicity $i \pi$ for the case
that the constraint (\ref{constraint}) is satisfied and for the
case treated in Section \ref{subsec:thetamp}, the $Q(u)$
functions have only $2 i \pi$ periodicity for the cases
treated in Sections \ref{subsec:alpham} and \ref{subsec:betam}.
(See Eqs. (\ref{Qold}), (\ref{Qtheta}), (\ref{Q}) and  (\ref{Qbeta}), 
respectively.) 

Two key steps in our approach for solving for the function $h(u)$
(which permits the recasting of the functional relations
(\ref{funcrltn}) as the vanishing of a determinant (\ref{det})) are
solving the quadratic equation (\ref{quadratic}) for $z(u)$, and
factoring the result, such as in (\ref{producth1}).  For the special cases
solved so far (namely, the case (\ref{constraint}) considered in
\cite{CLSW, Ne, NR}, and the new cases considered here), the
discriminants of the corresponding quadratic equations are perfect
squares, and the factorizations can be readily
carried out.  However, for general values of the boundary parameters,
the discriminant is no longer a perfect square; and factoring the
result becomes a formidable challenge.  Perhaps elliptic functions may
prove useful in this regard.  \footnote{An attempt along this line for
the case $p=1$ was considered in \cite{XX}.} We hope to report further
on this problem in the future.

\section*{Acknowledgments}

This work was supported in part by the National Science Foundation
under Grants PHY-0098088 and PHY-0244261.

\begin{appendix}

\section{Appendix}

Here we briefly review the solution \cite{Ne, NR} for the case that the
constraint (\ref{constraint}) is satisfied, in order to facilitate
comparison with the new cases considered in text. The matrix ${\cal M}$
is then given by
\be
{\cal M} = \left(
\begin{array}{cccccccc}
    \Lambda(u) & -h(-u-\eta) & 0  & \ldots  & 0 & -h(u)  \\
    -h(u+\eta) & \Lambda(u+\eta) & -h(-u-2\eta)  & \ldots  & 0 & 0  \\
    \vdots  & \vdots & \vdots & \ddots 
    & \vdots  & \vdots    \\
   -h(-u-(p+1)\eta)  & 0 & 0 & \ldots  & -h(u+p\eta) &
    \Lambda(u+p\eta)
\end{array} \right)  \,,
\label{calMold}
\ee
where $h(u)$ must satisfy
\be
h(u + i \pi) = h \left(u + (p+1)\eta \right) &=& h(u) \,,
\label{cond0old} \\
h(u+ \eta)\ h(-u-\eta) &=& \delta(u) \,, \label{cond1old} \\
\prod_{j=0}^{p} h(u+j\eta) + \prod_{j=0}^{p} h(-u-j\eta) &=& f(u) 
\,. \label{cond2old} 
\ee
A pair of solutions is given by $h(u)= h^{(\pm)}(u)= h_{0}(u) h_{1}^{(\pm)}(u)$ 
with $h_{0}(u)$ given by (\ref{h0}), and $h_{1}^{(\pm)}(u)$ given by
\be
h_{1}^{(\pm)}(u) = (-1)^{N+1} 4 \sinh(u \pm \alpha_{-}) \cosh(u \pm \beta_{-}) 
\sinh(u \pm \alpha_{+}) \cosh(u \pm \beta_{+}) \,,
\ee
Indeed, $h_{0}(u)$ satisfies
\be
h_{0}(u+ \eta)\ h_{0}(-u- \eta) &=& \delta_{0}(u) \,, \non \\
\prod_{j=0}^{p} h_{0}(u+ j\eta) = \prod_{j=0}^{p} h_{0}(-u-j\eta) &=&
f_{0}(u)  \,,
\ee
where $ \delta_{0}(u)$ is given by (\ref{delta0}), and $f_{0}(u)$ is given
by (\ref{f0}) and (\ref{f0odd}) for $p$ even and odd, respectively.
Moreover, $h_{1}^{(\pm)}(u)$ satisfies
\be
h_{1}^{(\pm)}(u+ \eta)\ h_{1}^{(\pm)}(-u- \eta) = \delta_{1}(u) \,, 
\ee
where $\delta_{1}(u)$ is given by (\ref{delta1}); and
\be 
\lefteqn{\prod_{j=0}^{p} h_{1}^{(\pm)}(u+ j\eta) 
+ \prod_{j=0}^{p} h_{1}^{(\pm)}(-u-j\eta) =
f_{1}(u)  - (-1)^{p(N+1)} 2^{1-2p}\sinh^{2}\left(2(p+1)u) \right)
\times}
\non \\
& &\times \left[ (-1)^{N} \cosh \left(
(p+1)(\alpha_{-}+\alpha_{+}+\beta_{-}+\beta_{+}) \right)
+\cosh\left( (p+1)(\theta_{-}-\theta_{+})\right) \right] 
\,, \label{rhs}
\ee
where $f_{1}(u)$ is given by (\ref{f1}) and (\ref{f1odd}) for $p$ even
and odd, respectively.  Hence, if the constraint (\ref{constraint}) is
satisfied, then the RHS of (\ref{rhs}) reduces to $f_{1}(u)$; hence,
all the conditions (\ref{cond0old})-(\ref{cond2old})
are fulfilled.  The corresponding expression for the transfer matrix
eigenvalues is given by
\be
\Lambda^{(\pm)}(u) = h^{(\pm)}(u) {Q^{(\pm)}(u-\eta)\over Q^{(\pm)}(u)} 
+ h^{(\pm)}(-u-\eta) {Q^{(\pm)}(u+\eta)\over Q^{(\pm)}(u)}  \,,
\ee
with
\be
Q^{(\pm)}(u) = \prod_{j=1}^{M^{(\pm)}} 
\sinh(u - u_{j}^{(\pm)}) \sinh(u + u_{j}^{(\pm)} + \eta) \,,
\qquad M^{(\pm)}={1\over 2}(N-1\pm k) \,,
\label{Qold}
\ee
and Bethe Ansatz equations
\be
{h^{(\pm)}(u_{j}^{(\pm)})\over h^{(\pm)}(-u_{j}^{(\pm)}-\eta)} = 
-{Q^{(\pm)}(u_{j}^{(\pm)}+\eta)\over Q^{(\pm)}(u_{j}^{(\pm)}-\eta)} \,, 
\qquad j = 1 \,, \ldots \,, M^{(\pm)} \,.
\ee

\end{appendix}

\vfill\eject 
 
\renewcommand{\theequation}{\arabic{equation}}
\setcounter{equation}{0}
\setcounter{footnote}{0}

\noindent    
{\large ADDENDUM to ``Bethe Ansatz derived from the functional
relations of the open XXZ chain for new special cases''}

\bigskip

In [1] (to which we refer hereafter by I), we obtain Bethe Ansatz
solutions for the transfer matrix eigenvalues of the open XXZ chain
for the special cases that the bulk anisotropy parameter has values
\be
\eta = {i \pi\over p+1}  \,, \qquad p= 2 \,, 4 \,, 6\,, \ldots \,,
\label{eta}
\ee 
and {\it one} of the boundary parameters 
$\{ \alpha_{-}, \alpha_{+},\beta_{-}, \beta_{+} \}$ is arbitrary,
and the remaining boundary parameters are zero. 
Here we show that those results can readily be extended to the cases
that any {\it two} of the boundary parameters 
$\{ \alpha_{-}, \alpha_{+},\beta_{-}, \beta_{+} \}$
are arbitrary and the remaining boundary parameters are either $\eta$ 
or $i \pi/2$. (We assume that $\theta_{-} = \theta_{+} \equiv \theta$.) For these
cases, the corresponding Hamiltonians have the conventional 
local form (see, e.g., [2])
\be
{\cal H }&=& \sum_{n=1}^{N-1} H_{n\,, n+1} 
+{1\over 2}\sinh \eta \Big[ 
\coth \alpha_{-} \tanh \beta_{-}\sigma_{1}^{z}
+ \csch \alpha_{-} \sech \beta_{-}\big( 
\cosh \theta\sigma_{1}^{x} 
+ i\sinh \theta\sigma_{1}^{y} \big) \non \\
&-& \coth \alpha_{+} \tanh \beta_{+} \sigma_{N}^{z}
+ \csch \alpha_{+} \sech \beta_{+}\big( 
\cosh \theta\sigma_{N}^{x}
+ i\sinh \theta\sigma_{N}^{y} \big)
\Big] \,,
\ee
where $H_{n\,, n+1}$ is given by (I3.23). The corresponding energy 
eigenvalues are related to the eigenvalues $\Lambda(u)$ of the 
transfer matrix $t(u)$ (I2.1) by
\be
E = c_{1} {\partial \over \partial u} \Lambda(u) \Big\vert_{u=0} 
+ c_{2} \,,
\ee
where
\be
c_{1} &=& -{1\over 16 \sinh \alpha_{-} \cosh \beta_{-}
\sinh \alpha_{+} \cosh \beta_{+} \sinh^{2N-1} \eta 
\cosh \eta} \,, \non \\
c_{2} &=& - {\sinh^{2}\eta  + N \cosh^{2}\eta\over 2 \cosh \eta} 
\,.
\ee 

\bigskip

{\Large \bf 1 \quad $\alpha_{-}\,, \alpha_{+}$ arbitrary}

\bigskip

For the case that $\alpha_{\pm}$ are arbitrary and $\beta_{\pm} = \eta$, we find that
\be
\sqrt{f_{1}(u)^{2} -4  \prod_{j=0}^{p} \delta_{1}\left(u+(2j-1)\eta\right)}
&=& 2^{-2 p+3} \cosh^{2}\left( (p+1)u \right) \sinh \left( (p+1)u 
\right) \non \\
&\times& 
\left[ \sinh \left( (p+1) \alpha_{-} \right) -(-1)^{N} \sinh \left( (p+1) 
\alpha_{+} \right) \right]
\,.
\ee
The key point is that the argument of the square root is a 
perfect square.
For definiteness, we henceforth restrict to {\it even} values of $N$.
It follows that the quantity $z(u)$ appearing in (I3.11) is now given by (cf. (I3.13))
\be
z(u) &=& 2^{-2(p-1)} \cosh^{2}\left( (p+1)u \right) 
\left[ \sinh \left( (p+1)u \right) \pm \sinh \left( (p+1) \alpha_{-} 
\right) \right] \non \\
&\times& \left[ \sinh \left( (p+1)u \right) \mp \sinh \left( (p+1) \alpha_{+} \right)
\right]\,.
\ee
Corresponding solutions of (I3.12) for $h_{1}(u)$ are (cf. (I3.14))
\be
h_{1}(u) = 4 \cosh^{2}(u -\eta) 
\sinh(u \mp \alpha_{-}) 
\sinh(u \pm \alpha_{+})
{\cosh\left({1\over 2}(u \pm \alpha_{-}+\eta) \right)\over 
 \cosh\left({1\over 2}(u \mp \alpha_{-}-\eta) \right)}
{\cosh\left({1\over 2}(u \mp \alpha_{+}+\eta) \right)\over 
 \cosh\left({1\over 2}(u \pm \alpha_{+}-\eta) \right)}
\,.
\ee
Hence, for $h(u) = h_{0}(u) h_{1}(u)$  we can take (cf. (I3.15))
\be
h(u) &=&  4\sinh^{2N}(u+\eta){\sinh(2u+2\eta)\over \sinh(2u+\eta)}
\cosh^{2}(u -\eta)   \non \\
&\times& \sinh(u-\alpha_{-}) \sinh(u+\alpha_{+})
{\cosh\left({1\over 2}(u+\alpha_{-}+\eta) \right)\over 
 \cosh\left({1\over 2}(u-\alpha_{-}-\eta) \right)} 
 {\cosh\left({1\over 2}(u-\alpha_{+}+\eta) \right)\over 
 \cosh\left({1\over 2}(u+\alpha_{+}-\eta) \right)}\,,
\ee
which indeed satisfies (I3.3)-(I3.5). The transfer matrix eigenvalues 
and Bethe Ansatz equations are given by (I3.17), (I3.18), (I3.21),
with (cf. (I3.20))
\be
M=N+2p+1 \,.
\ee

\bigskip

{\Large \bf 2 \quad $\beta_{-}\,, \beta_{+}$ arbitrary}

\bigskip

For the case that $\beta_{\pm}$ are arbitrary and $\alpha_{\pm} = \eta$, we find that
\be
\sqrt{f_{1}(u)^{2} -4  \prod_{j=0}^{p} \delta_{1}\left(u+(2j-1)\eta\right)}
&=& i 2^{-2 p+3} \sinh^{2}\left( (p+1)u \right) \cosh \left( (p+1)u 
\right) \non \\
&\times& 
\left[ \sinh \left( (p+1) \beta_{-} \right) - \sinh \left( (p+1) 
\beta_{+} \right) \right]
\,,
\ee
and therefore
\be
z(u) &=& 2^{-2(p-1)} \sinh^{2}\left( (p+1)u \right) 
\left[ \cosh \left( (p+1)u \right) \pm i \sinh \left( (p+1) \beta_{-} 
\right) \right] \non \\
&\times& \left[ \cosh \left( (p+1)u \right) \mp i\sinh \left( (p+1) 
\beta_{+} \right)
\right]\,.
\ee
Thus, we take the function $h(u)$ to be (cf. (I3.28))
\be
h(u) &=&  4\sinh^{2N}(u+\eta){\sinh(2u+2\eta)\over \sinh(2u+\eta)}
\sinh^{2}(u -\eta)   \non \\
&\times& \left( \cosh u +  i \sinh \beta_{-} \right) 
\left( \cosh u -  i \sinh \beta_{+} \right) 
\,,
\ee
which indeed satisfies (I3.25)-(I3.27). The transfer matrix eigenvalues 
and Bethe Ansatz equations are given by (I3.29), (I3.30), (I3.32),
with (cf. (I3.31))
\be
M=  N + 2p -1 \,.
\ee

\bigskip

{\Large \bf 3 \quad $\alpha_{-}\,, \beta_{-}$ arbitrary}

\bigskip

For the case that $\alpha_{-}\,, \beta_{-}$ are arbitrary and $\alpha_{+} = i \pi/2$,
$\beta_{+} = \eta$, we find that
\be
\sqrt{f_{1}(u)^{2} -4  \prod_{j=0}^{p} \delta_{1}\left(u+(2j-1)\eta\right)}
&=& 2^{-2 p+3} \cosh^{2}\left( (p+1)u \right) \sinh \left( (p+1)u 
\right) \non \\
&\times& 
\left[ \sinh \left( (p+1) \alpha_{-} \right) + (-1)^{p\over 2} i \cosh \left( (p+1) 
\beta_{-} \right) \right]
\,,
\ee
and therefore
\be
z(u) &=& 2^{-2(p-1)} \cosh^{2}\left( (p+1)u \right) 
\left[ \sinh \left( (p+1)u \right) \pm  \sinh \left( (p+1) \alpha_{-} 
\right) \right] \non \\
&\times& \left[ \sinh \left( (p+1)u \right) \pm (-1)^{p\over 2} i \cosh \left( (p+1) \beta_{-} 
\right) \right]\,.
\ee
For $h(u)$ we take
\be
h(u) &=&  4\sinh^{2N}(u+\eta){\sinh(2u+2\eta)\over \sinh(2u+\eta)}
\cosh (u -\eta) \cosh u  \non \\
&\times& \sinh(u-\alpha_{-}) 
{\cosh\left({1\over 2}(u+\alpha_{-}+\eta) \right)\over 
 \cosh\left({1\over 2}(u-\alpha_{-}-\eta) \right)} 
\left( \sinh u + i \cosh \beta_{-} \right)  
\,,
\ee
which satisfies (I3.3)-(I3.5). The transfer matrix eigenvalues 
and Bethe Ansatz equations are given by (I3.17), (I3.18), (I3.21),
with (cf. (I3.20))
\be
M=N+p \,.
\ee

Similar results hold for the case  that $\alpha_{+}\,, \beta_{+}$ are 
arbitrary and $\alpha_{-} = i \pi/2$, $\beta_{-} = \eta$, etc.

\section*{Acknowledgments}

This work was supported in part by the National Science Foundation
under Grant PHY-0244261.

\end{document}